\documentclass[12pt]{article} 

\makeatletter
\def\@normalsize{\@setsize\normalsize{10pt}\xpt\@xpt 
\abovedisplayskip 10pt plus2pt minus5pt\belowdisplayskip  
\abovedisplayskip \abovedisplayshortskip \z@ plus3pt\belowdisplayshortskip
6pt plus3pt minus3pt\let\@listi\@listI} 
\def\subsize{\@setsize\subsize{12pt}\xipt\@xipt}
\def\section{\@startsection
{section}{1}{\z@}{1.0ex plus 1ex minus .2ex}{.2ex plus .2ex}{\large\bf}} 
\def\subsection{\@startsection 
{subsection}{2}{\z@}{.2ex plus 1ex}{.2ex plus .2ex}{\subsize\bf}} 
\makeatother

\begin{document} 

\date{} 

\title{\Large\bf Multiple-Size Divide-and-Conquer
  Recurrences\thanks{Copyright \copyright\ 1997 by Ming-Yang Kao. A version
    of this work 
    appeared in {\it Proceedings 
      of the International Conference on Algorithms, the 1996 International
      Computer Symposium}, pages 159--161, National Sun Yat-Sen University,
    Kaohsiung, Taiwan, Republic of China, 1996. 
Reprinted in {\em {SIGACT News}}, 28(2):67--69, June
                 1997.}}

\author{Ming-Yang Kao\thanks{Research supported in part by NSF Grant
    CCR-9101385.} 
\\ Department of Computer Science
\\ Duke University
\\ Durham, NC 27708
\\ U.S.A.
\\ \
\\ kao@cs.duke.edu}

\maketitle 

\newtheorem{thm}{Theorem}

\thispagestyle{empty}

\newcommand{\ceiling}[1]{\lceil{#1}\rceil}

\section{Introduction}
This note reports a tight asymptotic solution to the following recurrence
on all positive integers $n$:
\begin{eqnarray}
  & T(n) = c{\cdot}n^\alpha{\cdot}\log^\beta{n} + \sum_{i=1}^{k}
  a_i{\cdot}T(\ceiling{b_i{\cdot}n}) & \mbox{for $n \geq n_0$,}
  \label{rec_1}
  \\ 
  & 0 < T(n) \leq d & \mbox{for $n < n_0$,} 
  \label{rec_2}
\end{eqnarray}
where
\begin{itemize}
\item $\alpha \geq 0, \beta \geq 0, c > 0, d > 0$,
\item $k$ is a positive integer,
\item $a_i > 0$ and $1 > b_i > 0$ for $i=1,\ldots,k$,
\item $n_0 \geq \max_{i=1}^k \frac{1}{1-b_i}$.
\end{itemize}
Since $n_0 \geq \max_{i=1}^k \frac{1}{1-b_i}$, $\ceiling{b_i{\cdot}n} \leq
n -1$ for all $b_i$ and $n \geq n_0$.  Thus, the $T(n)$ term on the
left-hand side of (\ref{rec_1}) is defined on $T$-terms with smaller $n$,
and (\ref{rec_2}) properly specifies the initial values of $T$.

A special case of this recurrence, namely, $k=1$, is discussed in
\cite{BHS80,Lueker80} and standard textbooks on algorithms and is used
extensively to analyze divide-and-conquer strategies \cite{AHU74,CLR91}.  A
specific recurrence with $k=2$ is used to analyze a divide-and-conquer
algorithm for selecting a key with a given rank \cite{AHU74,BFPRT73,CLR91}.

Let $g(x) = \sum_{i=1}^{k} a_i{\cdot}b_i^x$. The {\it characteristic
  equation} of the general recurrence is the equation $g(x) = 1$.  Our
solution to the general recurrence is summarized in the following theorem.
\begin{thm}
\label{the_thm}
If $r$ is the solution to the characteristic equation of the general
recurrence, then
\[T(n) = \left\{\begin{array}{ll}
    \Theta(n^r) & \mbox{if $r > \alpha$;}
\\
    \Theta(n^\alpha\log^{1+\beta}{n}) & \mbox{if $r = \alpha$;}
\\
    \Theta(n^\alpha\log^\beta{n}) & \mbox{if $r < \alpha$.}
\end{array}\right.
\]
\end{thm}

The key ingredient of our proof for this theorem is the notion of a
characteristic equation.  With this new notion, our proof is essentially
the same as that of the case with $k =1$ \cite{AHU74,BHS80,CLR91,Lueker80}.
This note concentrates on elaborating the characteristic equation's role in
our proof by detailing an upper bound proof for a certain case.  Once this
example is understood, it is straightforward to reconstruct a general proof
for Theorem~\ref{the_thm}.  Consequently, we omit the general proof for the
sake of brevity and clarity.

\section{An Example}
This section discusses the general recurrence with $k=3$.  To further focus
our attention on the characteristic equation's role, we assume that $\beta
= 0$, $r$ is a positive integer, and $r > \alpha$.  Then, according to
Theorem~\ref{the_thm}, $T(n)=\Theta(n^r)$.  We will only prove
$T(n)=O(n^r)$.  The lower bound proof is similar.

Let $S(n) = f_1{\cdot}n^r - f_2{\cdot}n^{r-\frac{1}{2}} -
f_3{\cdot}n^\alpha$.  It suffices to show that there exist some positive
constants $f_1, f_2, f_3$ such that $T(n) = O(S(n))$.  These constants and
some others are chosen as follows.  Without loss of generality, we assume
$b_1 < b_2 < b_3$.
\begin{eqnarray*}
  f_3 & = & \frac{c}{g(\alpha)-1}; 
  \\ f_2 & = & \mbox{any positive constant};
  \\ f_1 & = & f_2+f_3+1;
  \\ m_0 & = & \max \{n_0, \frac{1}{b_1},
  \left(\frac{f_1{\cdot}2^r{\cdot}\frac{1}{b_1}}
    {f_2{\cdot}\left(g(r-\frac{1}{2})-1\right)}\right)^2\};
  \\ M & = & \max_{n < m_0}\{1, T(n)\}.
\end{eqnarray*}
Note that since $0 < b_i < 1$ for all $b_i$, $g$ is a decreasing function.
Then since $r > \alpha$ and $r > r-\frac{1}{2}$, $g(\alpha) > 1$ and
$g(r-\frac{1}{2}) > 1$.  Thus, the above constants are all positive.  We
next consider the following new recurrence:
\begin{eqnarray}
  & R(n) = c{\cdot}n^\alpha + 
  a_1{\cdot}R(\ceiling{b_1{\cdot}n}) +
  a_2{\cdot}R(\ceiling{b_2{\cdot}n}) +
  a_3{\cdot}R(\ceiling{b_3{\cdot}n}) 
\label{r1}
& \mbox{for $n \geq m_0$,}
  \\ 
  &  R(n) = 1 & \mbox{for $n < m_0$,} 
\nonumber
\end{eqnarray}
It can be shown by induction that $T(n) \leq {M}{\cdot}R(n)$ for all $n$.
Thus, to prove $T(n) = O(S(n))$, it suffices to show $R(n) \leq S(n)$ for
all $n$.

{\it Base Case}: $R(m) \leq S(m)$ for all $m < m_0$.  This follows from the
choice of $f_1$.

Given some $n \geq m_0$, we need to show $R(n) \leq S(n)$.

{\it Induction Hypothesis}: $R(m) \leq S(m)$ for all integers $m$ where
$m_0 \leq m < n$.

{\it Induction Step}:
\begin{eqnarray}
  R(n) & \leq & 
  c{\cdot}n^\alpha + a_1{\cdot}S(\ceiling{b_1{\cdot}n}) +
  a_2{\cdot}S(\ceiling{b_2{\cdot}n}) + a_3{\cdot}S(\ceiling{b_3{\cdot}n}) 
\label{eqn_i1}
\\ & \leq & c{\cdot}n^\alpha + f_1{\cdot}g(r){\cdot}(n+\frac{1}{b_1})^r -
f_2{\cdot}g(r-\frac{1}{2}){\cdot}n^{r-\frac{1}{2}} -
f_3{\cdot}g(\alpha){\cdot}n^\alpha 
\label{eqn_i5}
\\ & \leq & c{\cdot}n^\alpha +
f_1{\cdot}g(r){\cdot}n^r+f_1{\cdot}2^r{\cdot}n^{r-1}{\cdot}\frac{1}{b_1} -
f_2{\cdot}g(r-\frac{1}{2}){\cdot}n^{r-\frac{1}{2}} -
f_3{\cdot}g(\alpha){\cdot}n^\alpha
\label{eqn_i6}
\end{eqnarray}
In this above derivation, 
\begin{itemize}
\item (\ref{eqn_i1}) follows from (\ref{r1}), the inequality $m_0 \geq
  n_0$, the base step and the induction hypothesis;
\item (\ref{eqn_i5}) follows from the fact that $\ceiling{b_i{\cdot}n} \leq
  b_i{\cdot}(n+\frac{1}{b_1})$;
\item (\ref{eqn_i6}) follows from the fact that $(n+\frac{1}{b_1})^r \leq
  n^r+2^r{\cdot}n^{r-1}{\cdot}\frac{1}{b_1}$ because $r$ is a positive
  integer and $m_0 \geq \frac{1}{b_1}$.
\end{itemize}
To finish the induction step, note that the right-hand side of
(\ref{eqn_i6}) is at most $S(n)$ as desired for the following reasons.
\begin{itemize}
\item By the choice of $f_3$, 
\( 
c{\cdot}n^\alpha + f_3{\cdot}g(\alpha){\cdot}n^\alpha \leq
-f_3{\cdot}n^\alpha. 
\)
\item 
Since $m_0 \geq   \left(\frac{f_1{\cdot}2^r{\cdot}\frac{1}{b_1}}
    {f_2{\cdot}\left(g(r-\frac{1}{2})-1\right)}\right)^2$, 
$
f_1{\cdot}2^r{\cdot}n^{r-1}{\cdot}\frac{1}{b_1} -
f_2{\cdot}g(r-\frac{1}{2}){\cdot}n^{r-\frac{1}{2}} \leq
- f_2{\cdot}n^{r-\frac{1}{2}}.$
\end{itemize}

\section*{Acknowledgments}
The author found the result in this note in 1986 while teaching a course on
algorithms. Since then, he has been teaching it in his classes.  He wishes
to thank Don Rose for helpful discussions.

%\bibliographystyle{plain} 
%\bibliography{all}

\end{document}